\documentclass[10pt,twocolumn,letterpaper]{article}

\usepackage{cvpr}              

\usepackage{fancyhdr}
\usepackage{comment}
\usepackage[ruled,vlined,linesnumbered]{algorithm2e}
\usepackage{amsmath}
\usepackage{mathtools}
\usepackage{booktabs} 
\usepackage{array}   
\usepackage{amsthm}

\usepackage{listings}
\usepackage{xcolor}

\definecolor{cvprblue}{rgb}{0.21,0.49,0.74}
\usepackage[pagebackref,breaklinks,colorlinks,allcolors=cvprblue]{hyperref}

\def\paperID{8672} 
\def\confName{CVPR}
\def\confYear{2025}

\title{ProReflow: Progressive Reflow with Decomposed Velocity }

\author{
Lei Ke\textsuperscript{1} \hspace{2mm}
Haohang Xu\textsuperscript{3} \hspace{2mm}
Xuefei Ning\textsuperscript{1} \hspace{2mm}
Yu Li\textsuperscript{1} \\
Jiajun Li\textsuperscript{4} \hspace{2mm}
Haoling Li\textsuperscript{1} \hspace{2mm}
Yuxuan Lin\textsuperscript{1} \hspace{2mm}
Dongsheng Jiang\textsuperscript{3} \\
Yujiu Yang\textsuperscript{1$\dagger$} \hspace{2mm}
Linfeng Zhang\textsuperscript{2$\dagger$} \\
\textsuperscript{1}Tsinghua University  \textsuperscript{2}Shanghai~Jiao~Tong~University \\
\textsuperscript{3}Huawei~Inc. \textsuperscript{4}University~of~Electronic~Science~and~Technology~of~China \\
\footnotesize 
\texttt{kl23@mails.tsinghua.edu.cn},  
\texttt{yang.yujiu@sz.tsinghua.edu.cn},  
\texttt{zhanglinfeng@sjtu.edu.cn}
}

\begin{document}
\maketitle
\renewcommand{\thefootnote}{}
\footnotetext[0]{$\dagger$Corresponding author.}
\begin{abstract}
    Diffusion models have achieved significant progress in both image and video generation while still suffering from huge computation costs. As an effective solution, flow matching aims to reflow the diffusion process of diffusion models into a straight line for a few-step and even one-step generation. However, in this paper, we suggest that the original training pipeline of flow matching is not optimal and introduce two techniques to improve it. Firstly, we introduce progressive reflow, which progressively reflows the diffusion models in local timesteps until the whole diffusion progresses, reducing the difficulty of flow matching. Second, we introduce aligned v-prediction, which highlights the importance of direction matching in flow matching over magnitude matching. Experimental results on SDv1.5 and SDXL demonstrate the effectiveness of our method, for example, conducting on SDv1.5 achieves an FID of 10.70 on MSCOCO2014 validation set with only 4 sampling steps, close to our teacher model (32 DDIM steps, FID = 10.05). Our codes will be released at Github.
\end{abstract}    
\section{Introduction}
\label{sec:intro}

Diffusion models have achieved significant breakthroughs in image and video generation, boosting various downstream applications such as text-to-image generation~\cite{ldm,peebles2023scalable} and image editing~\cite{couairon2022diffedit,dong2023prompt,kawar2023imagic}. However, compared with traditional generation models such as GANs~\cite{conditional_GAN}, the sampling process of diffusion models is formulated to include multiple timesteps, which severely harms its computation efficiency, hindering its application in edge devices and real-time applications. To solve this problem, abundant methods have been proposed to reduce the number of sampling steps such as step distillation~\cite{PD,meng2023distillation}, consistency models~\cite{CM,lcm_lora} and flow matching~\cite{flow_matching,reflow}.

\begin{figure}{}
  \centering
  \includegraphics[width=0.95\linewidth]{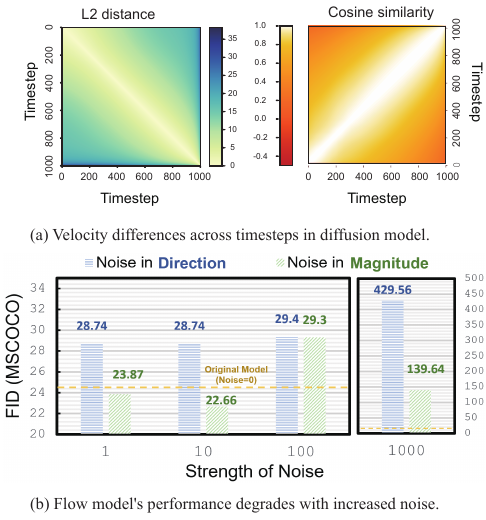}
  \vspace{-0.3cm}
  \caption{(a) L2 distance and Cosine similarity across velocities at different timesteps, the velocity discrepancy between timesteps increases with their distance in timesteps. (b) The consistently larger FID degradation under directional noise demonstrates that velocity direction is more critical for generation quality.}
  \vspace{-0.3cm}
  \label{fig:motivation}
  \vspace{-6pt}
\end{figure}

Among them, flow matching has gained popularity due to its simplicity and effectiveness. 
By re-flowing the pretrained diffusion models into a line, few steps and even 1-step generation can be achieved with tolerant loss in generation quality. The training process of reflow usually contains two steps. Firstly, the pretrained diffusion model generates abundant (noise, image) pairs. Then, the diffusion model is trained to make the velocity at different timesteps to be identical, indicating that the trajectory is rectified.
However, in this paper, we suggest that such a training strategy has not fully unleashed the potential of rectified flow and introduced two training techniques referred to as \emph{progressive reflow} and \emph{aligned v-prediction} to further improve it.

\noindent \paragraph{Progressive Reflow:} Traditional Reflow usually starts from a pretrained diffusion model and directly trains it to have a consistent prediction of velocity in all timesteps, which is theoretically correct but introduces difficulty in the optimization process.
As shown in Figure~\ref{fig:motivation} (a), the pretrained diffusion model has significantly different velocities at different timesteps, and directly eliminating these differences raises challenges in the training process. 

Fortunately, Figure~\ref{fig:motivation} (a) also shows that the pretrained diffusion model exhibits similar velocity in the adjacent timesteps, which provides the possibility to first reflow the model in a local window, and then reflow it in the whole training process. Such a progressive reflow pipeline allows the model to first learn to solve an easy problem and then extend to a difficult problem, which implies curriculum learning in generative models and thus facilitates the training process. Based on this observation, we propose progressive reflow, which firstly divides the whole diffusion process into $N$ windows, and then progressive reflow $N$ windows into $N/2$, $N/4$, $N/2$, $N/8$  until very few and even one window.

\noindent \textbf{Aligned V-Prediction:} Flow matching aims to match the velocity in different timesteps to achieve the target that the whole diffusion process is a straight line. However, we suggest that such a velocity matching is not optimal for the target of a ``straight line'' as the velocity can be further decomposed into its direction and magnitude, where the direction is more crucial for straightness. In other words, matching the direction of the velocity should have a higher priority than matching the magnitude, which has been ignored in previous works. Based on this observation, we propose to modify the original training loss of flow matching by introducing direction matching to solve this problem.

Our experiments have validated the effectiveness of the two improved training techniques. For instance, on MSCOCO-2017, 10.94 and 21.73 FID reduction can be observed with our ProReflow-II compared to rectified flow (2-ReFlow~\cite{instaflow}) at 4 steps and 2 steps respectively, demonstrating improvements in generation quality

In summary, our contributions are as follows.

\begin{itemize}
    \item We propose progressive reflow, which progressive reflow the diffusion model in local timesteps until the whole diffusion process. Progressive reflow implies the curriculum learning in flow matching and facilitates model training.
    \item Based on the observation that the direction of velocity is more crucial than the magnitude for straightness, we introduce velocity direction matching as an additional target for flow matching to facilitate model training.
    \item Extensive experiments demonstrate that both components are effective individually, and their combination achieves state-of-the-art performance with only 4 sampling steps.
\end{itemize}

\section{Related Work}
\label{sec:related work}

\begin{figure*}[!ht]
  \centering
  \includegraphics[width=0.95\linewidth]{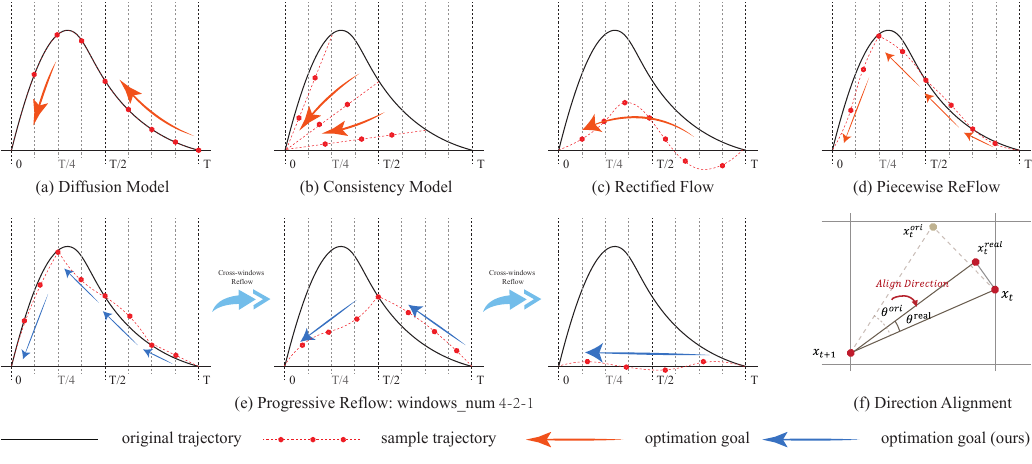}
  \caption{Conceptual illustration of different methods. (a)--(e) compare training objectives and sampling trajectories across different methods. Arrows show optimization targets, and red dashed lines represent actual sampling trajectories, which are curved due to the optimization not achieving the theoretical optimum. (e) shows our progressive reflow method achieves better approximation. (f) presents how our proposed aligned v-prediction works between timesteps $[t,t+1 ]$, it reduces prediction deviation with velocity direction correction.}
  \label{fig:progressive reflow}
\end{figure*}

\subsection{Text-to-Image Generation}

Diffusion models(DMs) learn the mapping from noise to images by fitting marginal probability distributions at each timestep~\cite{DDPM, score_based}. 
It works well because the forward diffusion process, which progressively adds noise to images, maintains the same marginal distributions as the sampling process~\cite{understanding}.
Combined with some technologies like classifier-free guidance and text encoder~\cite{CFG, GLIDE, Imagen}, DMs have surpassed GANs~\cite{GAN, StyleGAN} and VAEs~\cite{VAE, DALL-E} not only in generation quality but also in training stability.
Besides applied in pixel space, DMs can be effectively applied in latent space as well, which significantly reduces computational complexity~\cite{ldm}.
Despite achieving impressive generation quality, the iterative nature of DMs impacts its generation efficiency. Consequently, accelerating inference of diffusion models has emerged as an avtive research topic.

\subsection{Efficient Diffusion}

Existing approaches for accelerating DMs can be predominantly classified into two categories: efficient diffusion samplers and step distillation~\cite{DMD}.

The former category incorporates differential equation solvers into inference without requiring additional training.
DDIM~\cite{ddim} enables step skipping in the reverse process by introducing a non-Markovian sampling strategy.
DPM-Solver ~\cite{dpm-solver} reformulates the reverse diffusion process into an ODE system and solves it with high-order numerical methods, achieving superior sampling efficiency.
Sampler-based methods enable diffusion models to maintain satisfactory generation quality with 20 steps; however, performance deteriorates significantly when further reducing the step count (such as below 10).

The second category methods enhance few-step inference performance through another step distillation process. 
Progressive Distillation(PD)~\cite{PD} adopts a staged approach, iteratively halving the student model's sampling steps. 
Adversarial Diffusion Distillation(ADD)~\cite{ADD} leverages adversarial training for improved supervision, while Consistency Distillation (CD)~\cite{CM} enforces output convergence toward the target image across the sampling trajectory.

\subsection{Rectified Flow}

Flow matching has emerged as a kind of advanced diffusion model~\cite{flow_matching,reflow}.
It reformulates the forward process as a linear interpolation between noise and images, thereby proposing to predict a consistent velocity $v$ across the entire sampling trajectory. 
Thus, the sampling process is simplified to a temporal integration of the velocity field $v$.

Similarly, ReFlow was proposed as a technique applying flow matching to pretrained diffusion models, enabling the adaptation of existing architectures without retraining from scratch~\cite{reflow}.
InstaFlow~\cite{instaflow} first extended ReFlow to large-scale text-to-image models through consecutive ReFlow to straighten the ODE trajectory, followed by distillation to achieve single-step sampling.
Subsequently, some works explored improving ReFlow's effectiveness or simplifying its training\cite{accelerating, exploring, improving, minimizing}.
While ReFlow showed promise for single-step generation, its few-step sampling performance lagged behind state-of-the-art methods~\cite{pcm,hyper,PD}. To address this limitation, PeRFlow~\cite{perflow} proposed trajectory partitioning into time windows, achieving competitive few-step sampling through localized straightening within each temporal segment.

\subsection{Privileged Information in Distillation}
Although knowledge distillation has been proven effective as a model compression technique and further extended successfully to diffusion model acceleration, the theoretical explanation for its efficacy has remained elusive.
How 'dark knowledge' is effectively captured from teacher models and utilized to guide student learning remains a fundamental theoretical question~\cite{hinton2015distilling}.

Lopez-Paz \etal ~\cite{unifying} presented a unified theoretical framework that connects distillation with privileged information, establishing a generalized framework for understanding machine-to-machine knowledge transfer.
Viewing distillation as a transfer of privileged information, TAKD~\cite{TAKD} showed that an assistant model of intermediate capacity could more effectively mediate the knowledge flow between teacher and student models.

\section{Methods}
\label{sec:methods}

\begin{figure*}[t]
    \vspace{-20pt}
    \centering
    \begin{minipage}{0.47\textwidth}
        \centering
        \makebox[\textwidth]{\includegraphics[width=\textwidth]{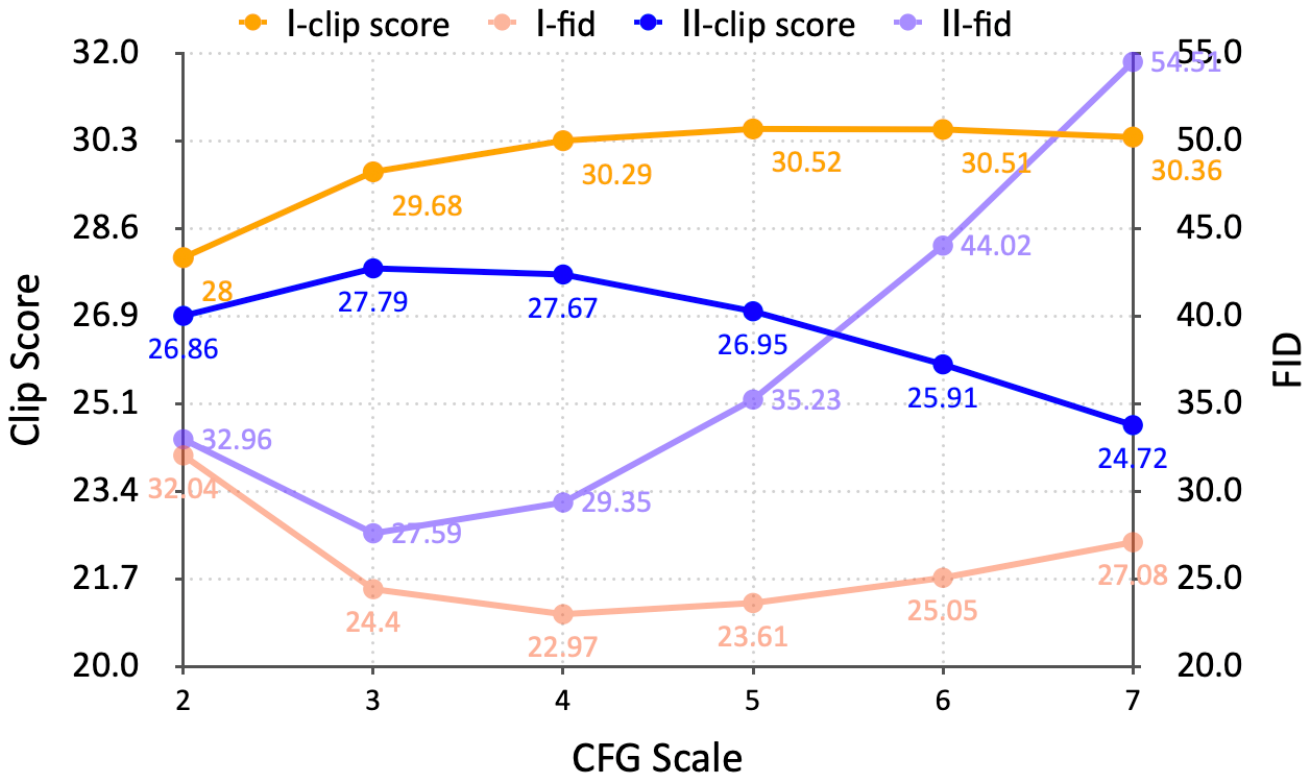}} 
        \vspace{-40pt}
        \caption{Performance of our models under different factors of classifier-free guidance (CFG) on COCO-2017. CFG scale ranges from 2 to 7. {\rm I} and {\rm II} stands for ProReflow-{\rm I} with 4 steps and ProReflow-{\rm II} with 2 steps, respectively.}
        \label{fig:cfg_fid_clip}
    \end{minipage}
    \hfill
    \begin{minipage}{0.47\textwidth}
        \centering
        \vspace{15pt}
        \makebox[\textwidth]
        {\includegraphics[width=0.7\textwidth]
        {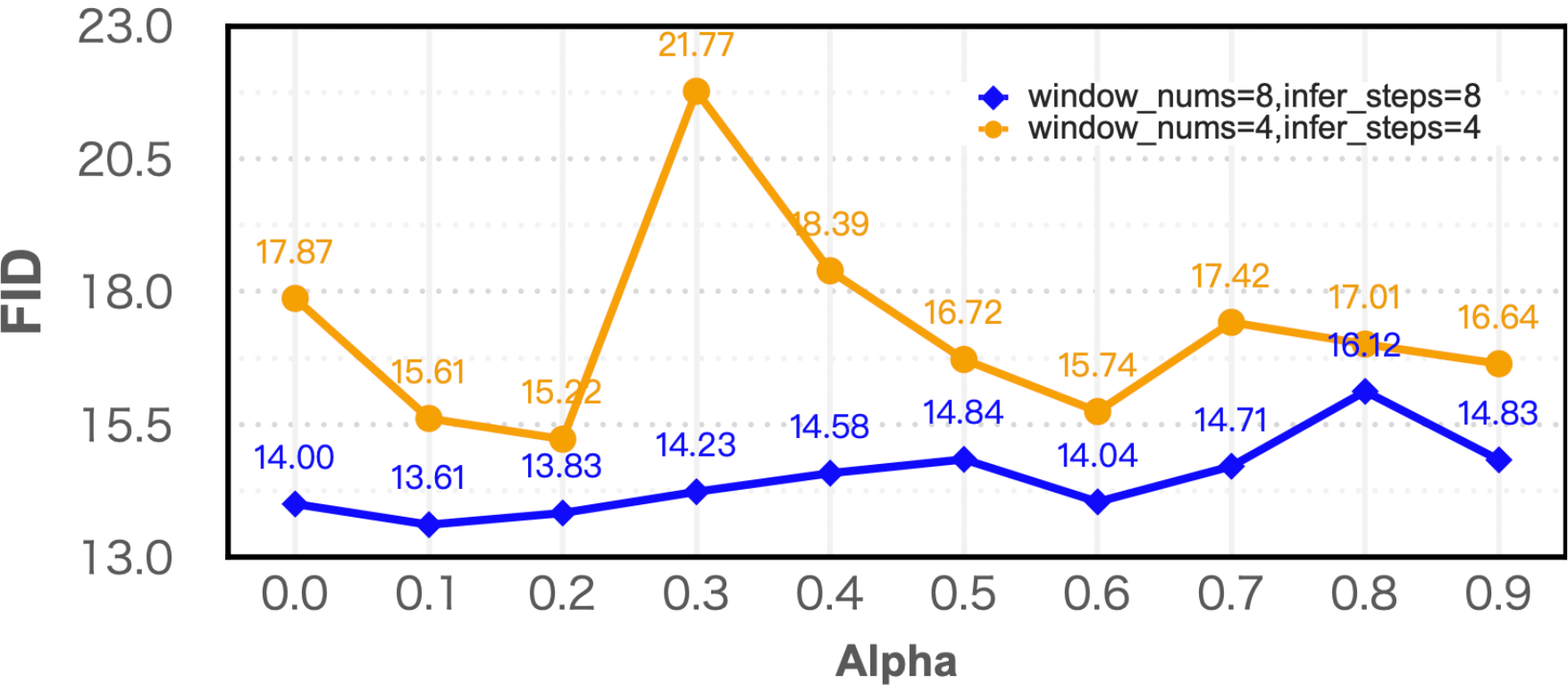}}
        \vspace{-30pt}
        \caption{FID on COCO-30K. The yellow curve shows results trained with 4 windows and evaluated using 4 inference steps, while the blue curve represents the model trained with 8 windows and evaluated using 8 inference steps. Both configurations are compared against their baselines where $\alpha=0$ (MSE loss only). Each model is trained for 10,000 iterations with batch size 32.}
        \label{fig:alpha_fid}
    \end{minipage}
    \label{fig:cfg_compare}
\end{figure*}

We present ProReflow, a more robust flow model training method. Our approach is motivated by the observation that training efficient few-step flow models faces two main challenges: (1) the significant trajectory approximation gap between teacher and student models, and (2) the difficulty in achieving accurate velocity prediction across large time intervals. 
To address these challenges, we propose progressive reflow for stable optimization of sample trajectory and aligned v-prediction for achieving precise velocity prediction, respectively, shown in Fig.~\ref{fig:progressive reflow}. 
\subsection{Temporal Segmentation for ReFlow}

{\bf Rectified Flow}
ReFlow aims to achieve temporally consistent velocity predictions across all timesteps. 
Given initial Gaussian distribution $\pi_0$ and target image distribution $\pi_1$, where $X_1 \sim \pi_1$ and $X_0 \sim \pi_0$.
Reflow defines a linear process from $\pi_0$ to $\pi_1$, where the corresponding sampling process follows the ODE:
\begin{equation}
    dX_t = v(X_t, t)dt, \quad t \in [0,1],
\label{eq:ode}
\end{equation}
Then, it formulates a least-squares optimization problem to ensure the predictions consistency:
\begin{equation}
    \min_{\theta} \int_0^1 \mathbb{E}[\|X_1 - X_0 - v_\theta(X_t, t)\|^2],
\label{eq:ori_loss}
\end{equation}
where $ X_t = tX_1 + (1-t)X_0 $.\\

\noindent{\bf Piecewise ReFlow} 
Aimed at improving the few-step generation, PeRFlow divides the sampling trajectory into multiple time windows, defined by endpoints $1 = t_K > \cdots > t_k > t_{k-1} > \cdots > t_0 = 0$.
Within each time window $[t_k, t_{k-1})$ formed by adjacent endpoints, PeRFlow assumes a linear process to straighten the trajectory, thus eq.\eqref{eq:ori_loss} can be reformulated as: 
\begin{equation}
   \min_{\theta} \sum_{k=1}^K \mathbb{E}_{z_{t_k} \sim \pi_k} \left[\int_{t_{k-1}}^{t_k} \|\frac{z_{t_{k-1}} - z_{t_k}}{t_{k-1} - t_k} - v_\theta(z_t, t)\|^2 dt\right],
   \label{eq:perflow}
\end{equation}
where $z_t = \alpha_t z_{t_k} + (1-\alpha_t) z_{t_{k-1}},\alpha_t = \frac{t - t_{k-1}}{t_k - t_{k-1}}$.
Finally, PeRFlow results in a piecewise linear trajectory composed of multiple segments.

\subsection{Progressive ReFlow}
PeRFlow originally sets the number of time windows to 4.
Despite achieving improvement in few-step inference, PeRFlow faces a significant optimization challenge: 
it attempts to approximate the teacher model's irregular trajectory using four linear intervals within a single training stage.

We propose a multi-stage progressive training scheme to tackle this challenge.
Rather than directly mapping the original trajectory to four time windows, our method first obtains an eight-window approximation from the original trajectory, and subsequently apply \emph{ Cross-windows ReFlow} to refine this eight-window representation into the target four-window configuration.

\noindent{\bf Cross-windows ReFlow} Consider three consecutive time points $t_{k-1}, t_k, t_{k+1}$. 
The optimization objectives in first training stage can be formulated as:
\begin{equation}
\begin{aligned}
    & \min_{\theta} \left( \mathbb{E}_{z_{t_k}\sim \pi_k} \int_{t_{k-1}}^{t_k} \| \frac{z_{t_{k-1}} - z_{t_k}}{t_{k-1} - t_k} - v_\theta(z_t, t)\|^2 dt \right. + \\
    & \left. \mathbb{E}_{z_{t_{k+1}}\sim \pi_{k+1}} \int_{t_k}^{t_{k+1}} \| \frac{z_{t_k} - z_{t_{k+1}}}{t_k - t_{k+1}} - v_\theta(z_t, t)\|^2 dt \right),
\end{aligned}
    \label{eq:two_segment}
\end{equation}
where $z_t = \begin{cases}
   \alpha_t z_{t_k} + (1-\alpha_t) z_{t_{k-1}}, & t \in [t_{k-1}, t_k) \\
   \beta_t z_{t_{k+1}} + (1-\beta_t) z_{t_k}, & t \in [t_k, t_{k+1})
\end{cases}$
with $\alpha_t = \frac{t - t_{k-1}}{t_k - t_{k-1}}$ and $\beta_t = \frac{t - t_k}{t_{k+1} - t_k}$.
In adjacent time windows, trajectories evolve from $z_{t_{k-1}}$ to $z_{t_k}$ in $[t_{k-1}, t_k]$, and from $z_{t_k}$ to $z_{t_{k+1}}$ in $[t_k, t_{k+1}]$. 

Cross-windows ReFlow aligns the optimization direction by guiding trajectories in both intervals to progress from $z_{t_{k-1}}$ towards $z_{t_{k+1}}$, thus eq.\eqref{eq:two_segment} can be reformulated as:
\begin{equation}
    \min_{\theta}  \mathbb{E}_{z_{t_{k+1}}\sim \pi_{k+1}} \int_{t_{k-1}}^{t_{k+1}} \| \frac{z_{t_{k-1}} - z_{t_{k+1}}}{t_{k-1} - t_{k+1}} - v_\theta(z_t, t)\|^2 dt ,
    \label{eq:cross_windows}
\end{equation}
where $z_t = \alpha_t z_{t_{k+1}} + (1-\alpha_t) z_{t_{k-1}},\alpha_t = \frac{t - t_{k-1}}{t_{k+1} - t_{k-1}}$.

\noindent{\bf Theoretical Explanation}
Based on the theoretical framework of knowledge distillation~\cite{unifying}, we can explain the effectiveness of Progressive ReFlow.
Consider three key functions: the teacher function $f_t \in \mathcal{F}_t$ representing the original diffusion trajectory, an intermediate function $f_a \in \mathcal{F}_a$ for the 8-segment approximation, and the student function $f_s \in \mathcal{F}_s$ for the target 4-segment representation.
According to the VC theory~\cite{VC_theory}, when the student learns directly from the teacher, the learning rate is bounded by:
\begin{equation}
    \mathcal{R}(f_s) - \mathcal{R}(f_t) \leq \mathcal{O}\left(\frac{|\mathcal{F}_s|_C}{n^\beta}\right) + \varepsilon_l,
\end{equation}
where $\beta \in [\frac{1}{2}, 1]$ denotes the learning rate associated with task difficulty, $\varepsilon_l$ represents the approximation error, $\mathcal{R}$ represents the error,  $\mathcal{O(\cdot)}$and $\mathcal{\epsilon}$ represent the estimation error and approximation error, respectively.. The challenge lies in the significant capacity gap between the complex trajectory and the 4-segment approximation, resulting in a small $\beta$ that indicates difficult learning.

Progressive ReFlow decomposes this challenging process into two stages:
\begin{align}
   \text{Stage 1: } & \mathcal{R}(f_a) - \mathcal{R}(f_t) \leq \mathcal{O}\left(\frac{|\mathcal{F}_a|_C}{n^{\beta_1}}\right) + \varepsilon_{at} ,\\
   \text{Stage 2: } & \mathcal{R}(f_s) - \mathcal{R}(f_a) \leq \mathcal{O}\left(\frac{|\mathcal{F}_s|_C}{n^{\beta_2}}\right) + \varepsilon_{sa} .
\end{align}
The effectiveness is theoretically guaranteed when:
\begin{equation}
   \mathcal{O}\left(\frac{|\mathcal{F}_a|_C}{n^{\beta_1}} + \frac{|\mathcal{F}_s|_C}{n^{\beta_2}}\right) + \varepsilon_{at} + \varepsilon_{sa} \leq \mathcal{O}\left(\frac{|\mathcal{F}_s|_C}{n^\beta}\right) + \varepsilon_s,
\end{equation}
this inequality is satisfied in practice due to two key principles: (1) 8-segment allows for better fitting of the teacher's complex sampling trajectory,leading to smaller combined approximation error ($\varepsilon_{at} + \varepsilon_{sa} < \varepsilon_l$), (2) Enhanced optimization efficiency through the progressive process, where each stage solves a simpler problem compared to direct optimization,resulting in $\beta_1, \beta_2 > \beta$.

\subsection{Aligned V-prediction} 


We analyzed approximate error in the optimization process and found that directional errors lead to more significant performance degradation compared to magnitude errors, shown in Fig.\ref{fig:motivation} (b).
We then propose \emph{aligned v-prediction}, which emphasizes direction alignment in training.

\noindent{\bf Direction Matters}
Consider two arbitrary points $z_{t_{i-1}}$ and $z_{t_i}$ along the trajectory.
Given the target vector $v = z_{t_i} - z_{t_{i-1}}$ and the model prediction $p$.
According to the law of cosines, the error between $v$ and $p$ can be expressed as:
\begin{equation}
r = |p|^2 + |v|^2 - 2|p||v|\cos \theta,
\end{equation}
where $\theta$ denotes the angle between $p$ and $v$. We analyze two extreme cases:

\noindent $\bullet$ Misaligned, accurate magnitude $\left(|p| = |v|, \theta = \epsilon \right)$: 
\begin{equation}
r_1 = 2|v|^2(1 - \cos \epsilon);
\end{equation}
$\bullet$ Aligned, inaccurate magnitude $\left( \theta = 0, |p| = |v| + \epsilon \right) $:
\begin{equation}
r_2 = (|v| + \epsilon)^2 + |v|^2 - 2|v|(|v| + \epsilon) = \epsilon^2.
\end{equation}
Let $y = r_1 - r_2$. Using Taylor expansion for small $\epsilon$:
\begin{equation}
y = (|v|^2 - 1)\epsilon^2.
\end{equation}
Our empirical measurements using real image-noise pairs during training show that $|v|$ typically ranges from 70 to 120, yielding $y > 0$ with a substantial margin. This indicates that directional errors lead to significantly larger performance degradation than magnitude errors.
\normalsize

\noindent{\bf Directional Alignment} Our analysis reveals that directional components of $v$ play a more crucial role in generation quality than magnitude.
Based on this, we proposed aligned v-prediction in flow matching, which incorporates directional alignment through cosine similarity measurements.
Specifically, we propose a novel flow matching loss function that places greater emphasis on directional alignment:
\begin{equation}
    L = (1-\alpha) \cdot \text{MSE}(v, pred) + \alpha \cdot (1 - \cos(v, pred)),
\end{equation}
where the first term provides basic magnitude consistency, the second term enforces explicit directional alignment via cosine similarity. 
The hyperparameter $\alpha$ balances the relative importance between magnitude and direction.

\begin{algorithm}[th]
\caption{ProReflow Algorithm}
\label{alg:ProReflow}
\SetAlgoLined
\KwIn{$\mathcal{D}$: dataset, $f_\phi$: teacher model, $K$: list of window numbers (e.g., [8,4,2]), $\alpha$: loss weight (default=0.1)}
Initialize student model $f_\theta \leftarrow f_\phi$\;
\For{$k$ in $K$}{
    Split time $[0,1]$ into $k$ windows\;
    \While{not converged}{
        Sample $x$ from dataset $\mathcal{D}$\;
        Sample $\epsilon \sim N(0,1)$\;
        Sample timestep $t$ and locate window $[t_1,t_2]$ s.t. $t \in [t_1,t_2]$\;
        $z_{t_2}=t_2*x+(1-t_2)\epsilon$\; 
        Compute $z_{t_1}$ using $f_\phi$\;
        $z_t=t*x+(1-t)\epsilon$\;
        Compute target velocity $v=x-\epsilon$\;
        Predict $v_\theta = f_\theta(z_t, t)$\;
        $\mathcal{L} = (1-\alpha)\text{MSE}(v, v_\theta) + \alpha\text{Dir}(v, v_\theta)$\;
        Update parameters $\theta$\;
    }
}
\KwOut{Trained model $f_\theta$}
\end{algorithm}

\noindent{\bf Hyperparameter Configuration}
Increasing the value of $\alpha$ enhances the directional supervision in the optimization objective.
When $\alpha = 0$, the loss function degenerates to the conventional MSE loss.
To determine the optimal hyperparameter configuration, we systematically evaluated different settings: We randomly sampled 0.8M images from LAION-art as our training set and fine-tuned SDv1.5 with different $\alpha$ values while maintaining the same number of windows.
We computed FID on coco-30k to evaluate these models.

As shown in Fig.~\ref{fig:alpha_fid}, the choice of $\alpha$ significantly impacts the model performance.
Among the evaluated $\alpha$ values, more positive gains were observed with windows = 4 compared to windows=8, which may be attributed to the increased importance of directional consistency at larger window spans. 
Our experiment results show that $\alpha$=0.1 works well in all experiment settings, thus we maintained $\alpha$=0.1 for subsequent experiments.

Combining \emph{progressive reflow} and \emph{aligned v-prediction}, we present ProReflow, as shown in Algorithm~\ref{alg:ProReflow}.

\section{Experiments}
\label{sec:experiments}

\subsection{Experiment Configuration} 
\noindent{\bf Model and Dataset}
We evaluate our proposed method primarily on Stable Diffusion v1.5 and Stable Diffusion XL.
During training, we freeze all modules except the UNet and employ BF16 mixed precision training.
For SDv1.5, we initialize our training process with windows numbers=8 and progressively apply our method to derive ProReflow-{\rm I} (4 windows), which subsequently serves as the basis for developing ProReflow-{\rm II}.
For SDXL, we adopt training configurations established in ProReflow-{\rm I} on SDXL to develop ProReflow-SDXL, achieving four-steps sampling.

As for multi-stage training, we maintain consistency in the teacher model's sampling trajectory across different training stages by fixing the total DDIM steps to 32. 
Specifically, when windows = 8, we use 4 DDIM steps within each window to derive the endpoint from the starting point. 
For windows = 4, we use 8 DDIM steps per window. 
This ensures that the teacher's sampling trajectory remains identical across different training stages, allowing for fair comparisons and stable optimization.

SDv1.5 is trained on the LAION-Art dataset, with all images center-cropped to $512 \times 512$ resolution following its default setup. For SDXL, we fine-tune the model using a combination of LAION-Art and 1.5 million samples from the laion2B-en-aesthetic dataset, with all images center-cropped to $1024 \times 1024$ resolution.
All experiments were conducted on 8 NVIDIA H20 GPUs.

\noindent{\bf Evaluation Setting}
Following common practice in text-to-image generation, we adopt two widely-used quantitative metrics: Fréchet Inception Distance (FID)~\cite{fid} and clip score~\cite{clip_score}.
The evaluation is mainly conducted on two standard benchmarks: MS COCO 2014 validation dataset~\cite{coco} and MS COCO 2017 validation dataset~\cite{coco}.

\begin{table}[t]
\centering
\caption{Performance comparison on COCO-2017 validation set, following the evaluation setup in ~\cite{rectified_diffusion}. Our method outperforms existing flow-based approaches.}
\begin{tabular*}{\columnwidth}{@{\extracolsep{\fill}}lccc@{}}
\toprule
Method   & Step & FID (↓) & CLIP Score(↑) \\
\midrule
2-Reflow~\cite{instaflow}    &   2   &   49.32   &  27.36  \\
2-Reflow~\cite{instaflow}    &   4   &   32.97   &  28.93  \\
Instaflow-0.9B~\cite{instaflow}    &   2   &   71.54   &  26.07  \\
Instaflow-0.9B~\cite{instaflow}    &   4   &   102.41  &  24.39  \\
PeRFlow~\cite{perflow}             &   4   &   23.81   &  30.24  \\
\textbf{ProReflow-{\rm I}} (ours)                  &   4   &   \textbf{22.97}   &  \textbf{30.29}  \\
\textbf{ProReflow-{\rm II}} (ours)                 &   2   &   27.59   &  27.79  \\
\textbf{ProReflow-{\rm II}} (ours)                &   4   &   \textbf{22.03}   &  29.95  \\
\bottomrule
\end{tabular*}
\label{tab:performance_coco5k}
\end{table}

\begin{table}[t]
\centering
\caption{Performance comparison on COCO-2014 validation set, following the evaluation setup in ~\cite{bk_sdm}.}
\begin{tabular*}{\columnwidth}{@{\extracolsep{\fill}}lccc@{}}
\toprule
Method & Time (↓) & Step & FID (↓)  \\
\midrule
\multicolumn{4}{l}{\textbf{ODE-solver based methods}} \\
DPMSolver~\cite{dpm-solver}       &  0.88s  &  25 &  9.78  \\
DPMSolver~\cite{dpm-solver}       &  0.34s  &  8  &  22.44 \\
DPMSolver++~\cite{dpm+solver++}   &  0.26s  &  4  &  22.36 \\
DDIM(our teacher)~\cite{ddim}                  &   $-$   &  32 &  10.05  \\
\midrule
\multicolumn{4}{l}{\textbf{Distillation-based methods}} \\
LCM-LoRA~\cite{lcm_lora}          &  0.12s  &  2  &  24.28  \\
LCM-LoRA~\cite{lcm_lora}          &  0.19s  &  4  &  23.62  \\
UniPC~\cite{unipc}                &  0.19s  &  4  &  23.30  \\
Flash Diffusion~\cite{flash_diffusion}          &  0.19s  &  4  &  12.41  \\
PCM~\cite{pcm}                    &  0.19s  &  4  &  11.70   \\

\midrule
\multicolumn{4}{l}{\textbf{Flow-based methods}} \\
Instaflow-0.9B~\cite{instaflow}   &   0.13s  &   2   &  24.61  \\
Instaflow-0.9B~\cite{instaflow}   &   0.21s  &   4   &  44.01  \\
2-ReFlow~\cite{instaflow}   &   0.13s  &   2   &  20.17  \\
2-ReFlow~\cite{instaflow}   &   0.21s  &   4   &  15.32  \\
PeRFlow~\cite{perflow}                           &   0.21s  &   4   &  12.01  \\
ProReflow-{\rm I} (ours)                &   0.21s  &   4   &  11.16  \\
ProReflow-{\rm II} (ours)                &   0.13s  &   2   &  15.44  \\
\textbf{ProReflow-{\rm II}} (ours)                &   0.21s  &   4   &  \textbf{10.70}  \\
\bottomrule
\end{tabular*}
\label{tab:performance_coco30k}
\end{table}

\subsection{Quantitative Results} 

We first compare our method with other flow-based acceleration approaches on COCO-2017 validation set, as shown in Table~\ref{tab:performance_coco5k}. With 4 inference steps, ProReflow-{\rm II} achieves an FID of 22.03 and a CLIP score of 29.95, showing significant improvements over 2-ReFlow, Instaflow and PeRFlow.Even with only 2 steps, ProReflow-{\rm II} maintains competitive performance. ProReflow-{\rm I} also demonstrates strong performance with an FID of 22.97 and the highest CLIP score of 30.29.
Table~\ref{tab:performance_coco30k} summarizes the comprehensive evaluation results Result on COCO-2014 valdation dataset with other diffusion acccelaration methods. With 32-step DDIM serving as our teacher model, ProReflow-{\rm II} achieves a competitive FID of 10.70 using only 4 steps. 

Table~\ref{tab:sdxl_results} presents a comprehensive comparison of our method with advanced acceleration approaches on SDXL. Our method achieves state-of-the-art performance while maintaining the same inference cost.

\begin{figure*}[t]
    \centering
    \includegraphics[width=\textwidth]{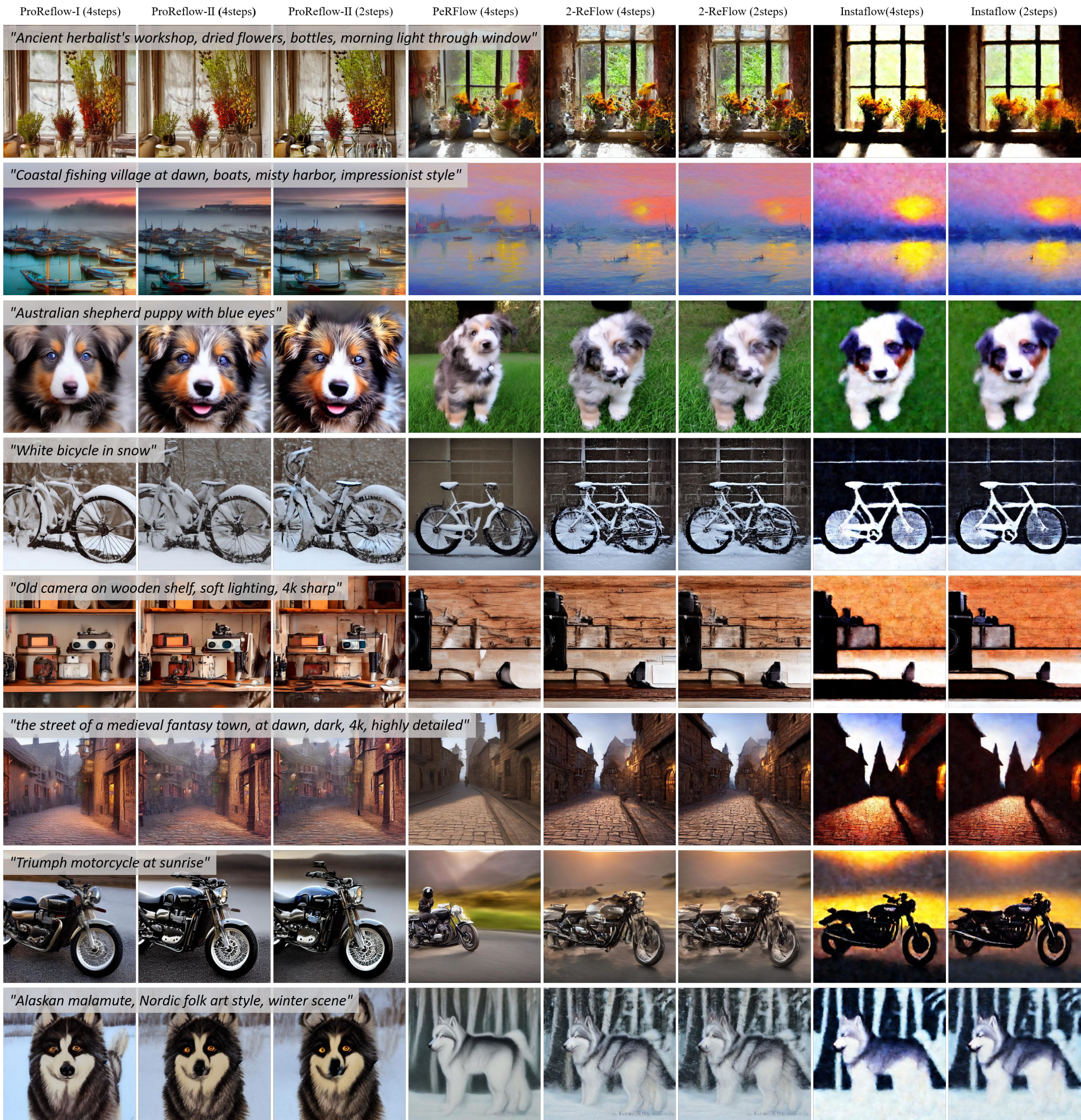}
    \caption{Qualitative comparison of image generation results. Our method demonstrates superior performance in detail rendering compared to other flow-based approaches at both 2-steps and 4-steps sampling.}
    \label{fig:show_case}
\end{figure*}

\begin{table}[t]
\centering
\caption{Comparison results on SDXL on COCO2017 validation set and COCO2014-10k validation set with 4 steps, following the evaluation setup in~\cite{rectified_diffusion}.}
\begin{tabular*}{\columnwidth}{@{\extracolsep{\fill}}lccc@{}}
\toprule
Method & Res. & Steps & FID (↓) \\
\midrule

\multicolumn{4}{l}{\textbf{COCO2017}} \\
Perflow          & 1024 & 4 & 27.06 \\
Rectified Diffusion & 1024 & 4 & 25.81 \\
\textbf{ProReflow-SDXL (Ours)} & 1024 & 4 & \textbf{25.36} \\

\midrule

\multicolumn{4}{l}{\textbf{COCO2014-10k}} \\
SDXL-Lightning   & 1024 & 4 & 24.56 \\
SDXL-Turbo       & 1024 & 4 & 23.19 \\
LCM              & 1024 & 4 & 22.16 \\
PCM              & 1024 & 4 & 21.04 \\
Perflow          & 1024 & 4 & 20.99 \\
Rectified Diffusion & 1024 & 4 & 19.71 \\
DMDv2            & 1024 & 4 & 19.32 \\
\textbf{ProReflow-SDXL (Ours)} & 1024 & 4 & \textbf{19.10} \\

\bottomrule
\end{tabular*}
\label{tab:sdxl_results}
\end{table}


\subsection{Qualitative Comparison} 
We compared our method against leading flow-based approaches (Rectified Flow, InstaFlow, and PerFlow) as shown in Figure \ref{fig:show_case}. Our method demonstrates superior performance across multiple aspects: it achieves more faithful detail preservation, renders more coherent global structures, and produces sharper textures with fewer artifacts. Specifically, while baseline methods often struggle with detail preservation and suffer from blurry regions or structural distortions, our approach consistently maintains both fine-grained details and global coherence across various scenarios. This comprehensive improvement in generation quality validates the effectiveness of our method.

\subsection{Training Cost} 

Although our method involves multiple training stages, its computational cost is significantly lower than 2-ReFlow, which applies ReFlow twice along the entire sampling trajectory and consumes 75.2 A100 days without considering data synthesis costs.
To obtain ProReflow-{\rm II} we perform three training stages starting from windows = 8, with each stage trained for 10000 iterations at a batch size of 256.
Despite the same total number of samples, the training time varies across stages.
Following~\cite{perflow}, for each batch we randomly sample a timestep and determine its corresponding window based on time windows division. 
The window's start and end points define the velocity prediction target, where starting points are obtained by directly adding noise to real images, and endpoints are generated by the teacher model. 
Since we maintain a total of 32 teacher inference steps across different stages, obtaining velocity targets for a batch with windows = 4 requires twice the teacher inference steps compared to windows = 8, which is consistent with the training time ratio between these stages.
Under this training framework, ProReflow-{\rm I} requires only 6.5 H20 days, and ProReflow-{\rm II} adds an additional 8.7 days, totaling 15.2 H20 days for the complete training pipeline.
Moreover according to NVIDIA's official specifications, the BF16 computation capability of H20 (148 TFLOPS) is approximately half of A100 (312 TFLOPS).

\section{Discussion}
\label{sec:discussion}
\subsection{Ablation Study}
We conduct ablation studies to examine our two core designs: \emph{aligned v-prediction} and \emph{progressive reflow}. Table~\ref{tab:ablation} presents results on COCO-2017 validation set. Both components contribute to model performance, with their combination yielding the best result.

\begin{table}[t]
\centering
\caption{Ablation studies on COCO-2017 validation set. We first show the results of gradually removing progressive reflow, aligned v-prediction, and both components, followed by our full model. We use a guidance scale of 4 for all the models.}
\begin{tabular*}{\columnwidth}{@{\extracolsep{\fill}}lccc@{}}
\toprule
Method & Steps & FID (↓) & CLIP (↑) \\
\midrule
w/o progressive reflow & 4 & 23.46 & 30.21 \\
w/o aligned v-prediction & 4 & 23.09 & 30.25 \\
w/o both & 4 & 23.81 & 30.24 \\
\textbf{ProReflow-{\rm I}} & 4 & \textbf{22.97} & \textbf{30.29} \\
\bottomrule
\end{tabular*}
\label{tab:ablation}
\end{table}

\subsection{CFG Influence}
It is well-established that the classifier-free guidance scale $w$ is a crucial factor affecting the performance of Stable Diffusion. During training, we set $w=1$ (i.e., without classifier-free guidance) throughout all the stages. To thoroughly understand the model's behavior under different guidance settings, we conducted extensive evaluations across a broad range of $w$ values from 2 to 7, measuring both FID and CLIP score, results are shown in Figure \ref{fig:cfg_fid_clip}.

\subsection{Step scalability}
Intuitively, for diffusion models, higher sampling steps should lead to better performance at the cost of increased inference time. However, this assumption does not always hold in practice. For instance, PeRFlow exhibits an unexpected performance degradation when increasing sampling steps from 4 to 8 on COCO-2014~\cite{perflow}, which limits its practical applications.
We surprisingly find our progressive training scheme effectively addresses this limitation. Although ProReflow-{\rm II} is trained with window size = 2, it achieves superior performance with 4-step sampling compared to ProReflow-{\rm I}, demonstrating both lower FID, shown in Table~\ref{tab:performance_coco5k} and Table~\ref{tab:performance_coco30k}.
\section{Conclusion}
\label{sec:limitation}
In this paper, we propose an efficient training framework for flow-based diffusion acceleration. If viewing the optimization process from temporal and spatial dimensions, our method naturally leads to two complementary techniques that correspond to these two dimensions respectively. Temporally, \emph{progressive reflow} bridges the trajectory approximation gap through curriculum learning, enabling gradual adaptation from more windows to fewer windows. Spatially, our velocity decomposition strategy emphasizes directional alignment over magnitude accuracy in velocity prediction. This principled design not only yields superior sampling quality but also brings advantages in optimization stability, training efficiency, and computational costs.

\noindent{\bf Limitations} Given promising few-step sampling performance, our method shows potential for one-step generation. However, due to computational constraints, we were unable to train the model with single window to full convergence. Nevertheless, we have validated the effectiveness of velocity decomposition in this challenging setting with the same training cost, only-one-window model equipped with aligned v-prediction demonstrate superior performance compared to the vanilla counterpart. We plan to move to one-step generation when resources allow.
{
    \small
    \bibliographystyle{ieeenat_fullname}
    \bibliography{main}
}

\clearpage
\setcounter{page}{1}
\maketitlesupplementary

\section{Velocity Gap in Long-Range Timesteps}
\label{sec:Velocity Gap}
As Fig.1 (a) of the main paper shown, to validate the velocity discrepancy in pretrained diffusion models, we conducted experiments using the Stable Diffusion v1.5 model. The velocity at each timestep is computed as:
\begin{equation}
v_t = 1000 \times (x_{t+1} - x_t)
\end{equation}
where $x_t$ represents the latent at timestep $t$. We sample 100 different prompts and average their velocity matrices to obtain reliable statistics. For each pair of timesteps $i$ and $j$, we compute both L2 distance $|V_i - V_j|_2$ and cosine similarity $\cos(V_i, V_j)$ between their velocities in the 4×64×64 latent space. All experiments use the PNDM scheduler with 1000 inference steps.

\section{Add noise to direction or magnitude}
\label{sec:add noise}
To analyze the relative importance of velocity direction versus magnitude in the flow model, we conduct experiments using the 2-Rectified model with 10 inference steps on COCO-5K validation set. For each velocity vector $v$, we decompose it into direction $d$ and magnitude $m$ components: $v = m \cdot d$, where $|d| = 1$.

For magnitude noise, we first add Gaussian noise to $m$ directly. Then, to ensure comparable perturbations for direction noise, we employ binary search to find an appropriate noise scale that yields the same L2 distance from the original velocity field as the magnitude noise. The directional noise is added to $d$ and then normalized to maintain unit length. This controlled noise injection mechanism enables fair comparison between directional and magnitude perturbations, with results shown in Fig.1 (b) of the main paper.

\begin{algorithm}[!h]
\caption{Velocity Decomposition}
\label{alg:VelocityDecomp}

\lstset{language=Python, frame=none, numbers=none}
\begin{lstlisting}
# Add directional constraint to standard MSE loss
def velocity_loss(v_pred, v_target):
    # Standard MSE loss
    l_mse = mse_loss(v_pred, v_target)
    
    # Additional directional constraint
    l_dir = 1 - cos_similarity(v_pred, v_target)
    
    # Weight between MSE and directional loss
    return (1-alpha)*l_mse + alpha*l_dir
\end{lstlisting}
\end{algorithm}

\section{ProReflow Implementary Details}
\label{sec:Implementary Details}
We have presented the pseudocode of ProReflow in Algorithm 1 in the main text. Here we elaborate on its two core components: Progressive ReFlow, which performs stage-wise training with decreasing window numbers [8,4,2], and the velocity decomposition loss which enhances directional alignment by incorporating cosine similarity alongside the standard MSE loss. The implementations are detailed in Algorithm \ref{alg:ProReflow-detail} and \ref{alg:VelocityDecomp}, respectively.

\begin{algorithm}[t]
\caption{Progressive ReFlow}
\label{alg:ProReflow-detail}

\lstset{language=Python, frame=none, numbers=none}
\begin{lstlisting}
# K: window numbers [8,4,2]
# D: training dataset
# t: normalized time in [0,1]

# Progressive window refinement
for windows in K:
    # Training in current stage
    while not converged:
        # Get endpoints of time window
        t = sample_time()  # t in [0,1]
        t1, t2 = get_window_bounds(t)
        
        # Compute trajectory endpoints
        z1 = add_noise(x0, t1)
        z2 = teacher_solve(z1, t1, t2)
        
        # Linear interpolation
        zt = interpolate(z1, z2, t)
        v_target = (z2 - z1)/(t2 - t1)
        
        # Update student model
        v_pred = student(zt, t)
        loss = velocity_loss(v_pred, v_target)
        update_params()
\end{lstlisting}
\end{algorithm}


\end{document}